%% file: main.tex
\documentclass[preprintnumbers,amsmath,amssymb,prd, notitlepage,nofootinbib,twocolumn, superscriptaddress]{revtex4}

\usepackage{graphicx}
\usepackage{dcolumn}
\usepackage{bm}
\usepackage{subfigure}
\usepackage{wasysym}
\usepackage{verbatim}
\usepackage{color}

\input{mydefinitions}

\usepackage{mathtools}
\usepackage{hyperref}
\usepackage{soul}

\newcommand{\rmn}{\mathrm}

\newcommand{\pwmap}{{\it Planck+WMAP }}
\newcommand{\lknee}{\ell_{\mathrm{knee}}}
\newcommand{\bEll}{{\boldsymbol{\ell}}}
\newcommand{\bL}{{\boldsymbol{L}}}
\newcommand{\bth}{\boldsymbol{\theta}}

\begin{document}

\title{Mitigating Foreground Biases in CMB Lensing Reconstruction Using Cleaned Gradients}

\author{Mathew S. Madhavacheril}
\affiliation{Department of Astrophysical Sciences, Princeton University, Princeton, NJ 08544, USA}
\email{mathewm@astro.princeton.edu}
\author{J. Colin Hill}
\affiliation{School of Natural Sciences, Institute for Advanced Study, Princeton, NJ 08540, USA}
\affiliation{Center for Computational Astrophysics, Flatiron Institute, New York, NY 10003, USA}

\date{\today}

\begin{abstract}
Reconstructed maps of the lensing convergence of the cosmic microwave background (CMB) will play a major role in precision cosmology in coming years. CMB lensing maps will enable calibration of the masses of high-redshift galaxy clusters and will yield precise measurements of the growth of cosmic structure through cross-correlations with galaxy surveys. During the next decade, CMB lensing reconstruction will rely heavily on temperature data, rather than polarization, thus necessitating a detailed understanding of biases due to extragalactic foregrounds.  In the near term, the most significant bias among these is that due to the thermal Sunyaev-Zel'dovich (tSZ) effect.  Moreover, high-resolution observations will be available at only a few frequencies, making full foreground cleaning challenging.  In this paper, we demonstrate a solution to the foreground bias problem that involves cleaning only the large-scale gradients of the CMB temperature map. We show that the data necessary for tSZ-bias-free CMB lensing maps already exist in the form of large-scale measurements of the CMB across multiple frequencies by the {\it Planck} and {\it WMAP} satellite experiments. Specifically, we show that the bias to halo masses inferred from CMB lensing is eliminated by the utilization of clean gradients obtained from multi-frequency component separation involving {\it Planck} and {\it WMAP} data, and that special lensing maps for galaxy cross-correlations can be prepared with only a small penalty in signal-to-noise while requiring no masking, in-painting, modeling, or simulation effort for the tSZ bias. While we focus on cross-correlations, we also show that gradient cleaning can mitigate biases to the CMB lensing autospectrum that arise from the presence of foregrounds in temperature and polarization with minimal loss of signal-to-noise.
\end{abstract}
\maketitle

\section{Introduction} \label{sec:Intro}

Gravitational lensing of the cosmic microwave background (CMB) has emerged as a powerful probe of the late-time matter distribution (see, e.g., \cite{lewisChallinor} for a review). Over the past decade, the first detections of CMB lensing both in cross-correlation \cite{ksmithlensing,hirata2008} and auto-correlation \cite{daslensing,engelenlensing} have progressed to a 2.5\% measurement of the amplitude of the power spectrum of the CMB lensing convergence field by {\it Planck} \cite{plancklensing2013,plancklensing2015}, with ground-based measurements rapidly approaching similar levels of sensitivity while providing information on smaller scales \cite{sherwin2015,storylensing,omorilensing,pblensing1,pblensing2,biceplensing}. The CMB lensing auto-spectrum is sensitive to the amplitude of structure formation at redshifts $z \approx 1-2$ and consequently provides constraints on the sum of the neutrino masses and the dark energy equation of state when combined with measurements of the primary CMB \cite{Allison2015,sherwin2015,simardlensing}. 

Maps of the CMB lensing convergence field will be a key ingredient in CMB polarization experiments that aim to detect primordial gravitational waves via their B-mode polarization signature. Lensing maps can be used to subtract the lensing contribution to observed B-mode maps, significantly improving constraints on the tensor-to-scalar ratio $r$ \cite{delensing1,delensing2,cmbs4,wudesigning,yu2017}. CMB lensing maps can also be cross-correlated with low-redshift large-scale structure tracers (e.g. \cite{plancklensing2013,Giannantonio,omoriHolder,baxterXcorr,miyatakeDR,tSZxCMBlensing}) and galaxy weak lensing maps (e.g. \cite{handShear,liuPlanck,kirk,kidsShear}). Such cross-correlations provide an alternate channel for constraining cosmological parameters like the amplitude of structure formation, neutrino mass, and the dark energy equation of state, and have different (and often less problematic) systematics than those in auto-spectrum measurements. They can be used to measure the bias of tracers and consequently primordial non-Gaussianity ($f_{\rmn{NL}}$) through its scale-dependent bias signature \cite{lensingfnl}. Cross-correlations also allow for the calibration of galaxy shear estimators and the photometric redshift distributions of galaxies in optical surveys \cite{Vallinotto,cmbShear,liuShear,schaanXcorr,baxterXcorr,miyatakeDR}.

Recently, CMB surveys have achieved sufficient resolution and sensitivity to detect the one-halo contribution to the CMB lensing signal from dark matter halos \cite{Baxter2015,Mat2015,PlnkSZCos2015,Baxter2017,geachpeacock}. This ``halo lensing'' signal provides an avenue for calibrating the masses of galaxy clusters, a crucial ingredient in cosmological analyses that involve cluster abundances. CMB halo lensing is powerful at high redshifts $z>1$ where optical galaxy surveys have both low statistics due to the steep decline in galaxy counts and suffer from systematics such as catastrophic photometric redshift failures \cite{lewisking,mat2017}. CMB halo lensing-calibrated galaxy clusters can provide constraints on cosmological parameters like the sum of the neutrino masses that are comparable and complementary to those obtained from the CMB lensing auto-spectrum~\cite{mat2017,LA2017}.

Apart from lensing, other secondary anisotropies sourced by late-time astrophysical processes are present in CMB maps. These secondaries can introduce biases in both CMB lensing maps and measurements of the lensing power spectrum~\cite{vanengelen,osborne2014,kszbias,Baxter2018}. The thermal Sunyaev-Zel'dovich (tSZ) effect due to Compton up-scattering of CMB photons off hot, ionized gas is one such foreground~\cite{SZ1969,SZ1970}. Biases due to the tSZ foreground were explored and quantified through simulations in \cite{vanengelen}. Very recently, \cite{Baxter2018} studied the problem of tSZ-induced biases in CMB lensing cross-correlations using a data-driven approach, advocating for aggressive masking and additional angular scale cuts as mitigation methods, at the cost of a significant decrease in signal-to-noise (S/N).

In this paper, we examine biases to CMB lensing cross-correlations and CMB halo lensing measurements using the same simulations \cite{Sehgal2011} as used in \cite{vanengelen}. We explore a solution to these biases (first suggested in \cite{mat2017}) that utilizes existing data from the {\it Planck} and {\it WMAP} experiments to obtain foreground-cleaned gradients of the CMB.  We show that cleaned gradient maps are sufficient to eliminate foreground-induced biases without a significant S/N penalty. We focus on biases due to the tSZ effect since these are known to be the most significant for low-redshift cross-correlations. However, our findings can be directly generalized to other foreground contaminants that can be removed using multifrequency data, e.g., the cosmic infrared background (CIB).

We begin by setting up the framework for ``asymmetric lensing reconstruction'' in Section \ref{sec:AsymLens}, i.e., lensing reconstruction where CMB gradients are obtained from a different experiment or use different frequency channel combinations than the non-gradient part. In Section~\ref{sec:ForeBias}, we motivate this framework further by examining how astrophysical foregrounds bias CMB lensing reconstruction and pointing out how cleaned gradients are sufficient for bias-free cross-correlations and halo mass inference. We survey existing measurements by {\it Planck} and {\it WMAP} and contrast between the SMICA foreground-cleaned {\it Planck} map and the LGMCA foreground-cleaned \pwmap map \cite{LGMCA1,LGMCA2}. By stacking on SDSS DR8 redMaPPer galaxy clusters~\cite{Redmapper}, we motivate the LGMCA map as our source of tSZ-free CMB gradients. Next, in Section~\ref{sec:Sims}, we explicitly perform CMB lensing reconstruction on realistic simulations from \cite{Sehgal2011} and demonstrate that cross-correlations with halo positions and reconstructions of halo masses are both bias-free when utilizing cleaned gradients. In Section~\ref{sec:SN}, we focus on forecasts that show that there is negligible loss of S/N when using LGMCA gradients in place of potentially contaminated high-resolution maps for the gradient. We conclude in Section~\ref{sec:Conclusion} with a discussion that includes applications of gradient cleaning to measurements of the CMB lensing auto-spectrum.

\section{Asymmetric lensing reconstruction}
\label{sec:AsymLens}

Gravitational lensing remaps points on the primary CMB sky while conserving surface brightness. In Fourier space, this corresponds to a previously Gaussian random field with independent Fourier modes developing mode-coupling that is proportional to the projected lensing potential~(e.g.,~\cite{Zaldarriaga1998,Zaldarriaga1999,huok}). For example, for the CMB temperature field (in the flat-sky approximation),

\begin{equation}
\langle T(\bEll) T(\bEll') \rangle_{{\rmn{CMB}}} = f^{TT}(\bEll,\bEll') \phi(\bL)
\end{equation}
where $\bEll+\bEll'=\bL$ with $\bL\neq 0$,  $f^{TT}(\bEll,\bEll')$ is a response function that depends on the CMB power spectra~\cite{huok}, $T(\bEll)$ is the Fourier transform of the CMB temperature field, $\phi(\bL)$ is the Fourier transform of the line-of-sight-projected gravitational potential, and the expectation value is over realizations of the unlensed CMB.

This intuition allows one to write a quadratic estimator to reconstruct the lensing mass distribution that sums over pairs of CMB modes with weights designed to minimize the variance of the reconstruction and normalization ensuring that the estimator is unbiased \cite{huok}.  As its name implies, the quadratic estimator requires a pair of maps. These pairs can be combinations of temperature (T), E-mode (E), or B-mode (B) polarization maps, allowing for estimator combinations of the form TT, TE, ET, EE, EB, and TB. In the absence of instrumental noise, the combination EB is the most constraining estimator across all scales since (in the absence of primordial B-modes) any fluctuation in the B-mode map comes from primordial E-modes that have been lensed into B-mode form.  For the noise levels of current data, the combination TT is the most constraining~\cite{plancklensing2015,omorilensing,sherwin2015}.  In practice, whether EB is better than TT depends both on the instrumental noise and the lensing wavenumber $L$ under question, with EB becoming comparable to TT for low wavenumbers at a noise level of around $4$ $\mu$K-arcmin and at noise levels much lower than $1$ $\mu$K-arcmin for the high wavenumbers relevant for clusters.

The quadratic estimator can be written as a product of filtered real-space fields~\cite{HDV2007}, which allows for efficient computation. For $X \in \{T,E,B\}$ and $Y \in \{T,E,B\}$
\begin{equation}
\label{Eq:kappaEst}
\kappa^{XY}(\boldsymbol{\theta}) =-\mathcal{F}^{-1}\left\{A^{XY}(\bL) \mathcal{F}\left\{ \rmn{Re}\left[\nabla \cdot \left[\boldsymbol{\nabla}X_f(\boldsymbol{\theta}) Y_f(\boldsymbol{\theta})^* \right]\right]\right\}\right\}
\end{equation}
where the subscript $f$ denotes filtering\footnote{The implementation used in this paper can be found in the \texttt{lensing} module of the open-source code \url{https://www.github.com/msyriac/orphics}}, $A(\boldsymbol{L})$ is a normalization in Fourier space that ensures this estimator is unbiased as a function of angular wavevector $\boldsymbol{L}$, $\kappa = -\frac{1}{2} \nabla^2 \phi$ is the reconstructed CMB lensing convergence, and $\mathcal{F}$ and $\mathcal{F}^{-1}$ represent 2D Fourier and inverse-Fourier transforms respectively. We refer the reader to \cite{HDV2007} for details of the optimal filtering and the form of the normalization $A(\boldsymbol{L})$.  We note that the above expression is not the usual symmetrized form of the estimator \cite{huok}. Even when only a single experiment is involved in mapping T, E, and B, the ET and TE estimators as written above contain some independent information and are not 100\% correlated.

We now call particular emphasis to the fact that the $X$ and $Y$ maps can be from different experiments, can involve combinations of maps observed over different frequency bandpasses, or can use different treatments of the non-CMB fluctuations. We will argue that for the purposes of cross-correlations with low-redshift tracers and for CMB lensing-inferred halo masses, lensing maps should be reconstructed in such a way that the ``gradient leg'' is ensured to be free of tSZ (if not all foregrounds). Once we allow for the filtering and noise in the two legs to be different, the estimator for TT is of course no longer symmetric. While no new information would have been gained if one used both $\kappa^{XY}$ and $\kappa^{YX}$ when the experiments used to measure $X=T$ and $Y=T$ are the same, $\kappa^{XY}$ and $\kappa^{YX}$ will have some independent information even for the TT estimator when the gradient is measured using a different experiment or set of frequencies.   In particular, using a low-resolution experiment (like {\it Planck}) for the temperature gradient introduces a far lower penalty in S/N than if such an experiment were used for the non-gradient leg. As pointed out in {\cite{HDV2007}}, this happens because the RMS gradient of the unlensed CMB saturates by around $\ell=2000$ due to Silk damping of the CMB.  Thus, the information gain in measuring the gradient more precisely slows down near that scale, which is also close to where {\it Planck} becomes noise-dominated. This observation will be key for the rest of this paper.

The information loss in using low-resolution gradients becomes even more negligible in the case of halo lensing. In the squeezed limit where we are interested in $\phi(\bL)$ for $L\rightarrow \infty$, the quadratic estimator pairs that contribute most to the S/N are precisely the couplings of the gradient on the largest scales and lens-induced fluctuations on the smallest scales.  Physically, when the mass distribution is small in angular size and azimuthally symmetric, the halo lensing signal is a local reduction of the unlensed background gradient resulting in a residual that is a dipole anti-aligned with the background gradient (see Figure 1 of \cite{lewisking}). We thus look for the correlation of the true, unlensed background gradient and the small-scale dipole. In fact, the imposition of a low-pass filter for the gradient (at $\ell_G = 2000$) is necessary to avoid a bias in the halo mass inference due to the halo itself affecting the quadratic estimator's knowledge of the ``true unlensed background gradient''~\cite{HDV2007}. Of course, this true gradient is not observable, and so there will always be some bias. This bias is mass-dependent, since more massive halos will affect the gradient more. The bias is around $2\%$ for a halo of mass $M_{200} = 2\times 10^{14} M_{\odot}$ and can in principle be calibrated out since its origin is well-understood. Alternative techniques are discussed in Section~\ref{sec:Conclusion}.

In the rest of this paper, we will focus on two applications of lensing reconstruction where the insights of asymmetric lensing reconstruction are particularly useful. The first is the inference of halo masses, where the procedure is generally to reconstruct lensing using quadratic estimators centered on halos and then stack many halos to average down reconstruction noise.  The second is the cross-correlation of a CMB lensing map with an external low-redshift tracer, e.g., galaxy overdensities, or with galaxy lensing maps. We also briefly discuss applications to measurements of the CMB lensing auto-spectra in Section~\ref{sec:Conclusion}. 

\section{Foreground Biases}
\label{sec:ForeBias}
We now consider CMB lensing reconstruction in the presence of astrophysical foregrounds. Observations by ground-based, high-resolution experiments will be vital in improving our measurements of CMB lensing over the next decade. A large fraction of the sensitivity to lensing will come from temperature maps, at least until these maps become deeper than $4$ $\mu$K-arcmin for large-scale CMB lensing and deeper than $0.1$ $\mu$K-arcmin for CMB halo lensing. At the frequencies of interest, a number of astrophysical foregrounds are present in CMB temperature maps, including the tSZ effect (Compton up-scattering of CMB photons by hot, ionized gas, producing a unique spectral distortion), emission from dusty, star-forming galaxies (i.e., the CIB), and the kinematic Sunyaev-Zel'dovich (kSZ) effect (Doppler boosting of CMB photons due to Compton scattering off electrons with non-zero line-of sight velocity, an effect which has the same frequency dependence as the primary CMB, to lowest order).

One can combine mm-wave maps at multiple frequencies (e.g., via the internal linear combination (ILC) method~\cite{WMAP1,Eriksen2004}) to preserve the CMB signal\footnote{Note that the kSZ and ISW signals are also preserved.} while removing non-blackbody foreground signals (e.g., tSZ and CIB). However, as has been noted earlier and is discussed again later in this paper, such cleaning methods are not perfect and can leave residuals in the final map.  Moreover, the choice of analysis method can strongly affect such residuals. If we know the frequency dependence of a foreground component (e.g., the tSZ spectral function), we can explicitly deproject it under the constrained ILC formalism (CILC)~\cite{Remazeilles2011}, thereby strongly suppressing residuals associated with this contaminant in the final map. This constraint comes with the penalty of increased noise in the final map, as the degree of freedom used for the deprojection is no longer available for variance-minimization. Nevertheless, as we show later, such deprojections can be extremely useful for some applications, e.g., asymmetric lensing reconstruction.  Finally, we note that before a survey is fully completed, one may not have deep measurements across the range of frequencies required to perform CILC (indeed, most ground-based CMB experiments to date have been limited to 2-3 frequency channels, although this will change in the coming years). 

Given this context, we consider the question of how foreground biases enter lensing and how we might avoid them. The real-space rewriting of the quadratic estimator provides useful insight. For simplicity, we consider the unnormalized estimator $\tilde{\kappa}(\bth)$ defined via $\kappa(\bth) = -\mathcal{F}^{-1}\left\{A^{XY}(\bL) \mathcal{F}\left\{\tilde{\kappa}(\bth)\right\}\right\} $ from Equation~\ref{Eq:kappaEst}, and we only consider the TT estimator. The observed field $T^o(\bth)$ contains both the lensed beam-convolved temperature $T(\bth)$ and the beam-convolved foreground field $F(\bth)$. We have
\begin{align}
  \langle\tilde{\kappa}(\bth)\rangle & = \langle\nabla \cdot \left[[\boldsymbol{\nabla}T^o(\bth)]_f T^o_f(\bth) \right]\rangle & \\
   & = \langle\nabla \cdot \left[[\boldsymbol{\nabla}T_f(\bth)+\boldsymbol{\nabla}F_f(\bth)] [T_f(\bth)+F_f(\bth)] \right]\rangle & \\
   & = \langle\nabla \cdot \left[[\boldsymbol{\nabla}T_f(\bth)] T_f(\bth)) \right]\rangle + \langle\nabla \cdot \left[[\boldsymbol{\nabla}F_f(\bth)] F_f(\bth) \right]\rangle
\label{Eq:kappaFore}
\end{align}
where in going to the last line we have assumed that, on average, gradients of either the CMB or the foreground do not correlate with the foreground or the CMB itself and vice versa. This is easy to see in the special case of an azimuthally symmetric foreground centered on a halo. Nevertheless, we will explicitly check with simulations for both the large-scale lensing case and the halo case that this assumption holds.

The second term in Equation~\ref{Eq:kappaFore} is a bias term. For large-scale lensing cross-correlations, the expectation value should be taken after correlating with the external tracer, in which case the bias from the second term appears as a bispectrum.  Crucially, we see that this bias term is zero if the gradient maps were free of foregrounds to start with. Hence, for bias-free halo masses and cross-correlations, it is sufficient to ensure the gradient is free of foregrounds, regardless of whether the non-gradient leg is foreground-contaminated.\footnote{Of course, foregrounds in the non-gradient leg still contribute variance to the final reconstruction. Standard ILC (or other component separation methods) without explicit deprojection can be applied using the available frequencies to reduce this variance.}

\subsection{tSZ bias}

We now specialize to the case of biases from the tSZ effect. The CMB lensing kernel has significant overlap with large-scale structure at low redshifts. This allows one to cross-correlate CMB lensing maps with low-redshift tracers, e.g., galaxies from an optical survey, and also with lensing shear maps as measured using background galaxies. From such cross-correlations, a variety of cosmological information can be gleaned, including the bias of tracers (and consequently primordial non-Gaussianity through large-scale scale-dependent bias), and the amplitude of fluctuations as a function of redshift (and consequently the sum of the neutrino masses). In addition, the amplitude of these cross-correlations can be used to constrain galaxy shear multiplicative bias and the photometric redshift distribution of optical surveys.

However, any tSZ contamination present in the temperature maps used for CMB lensing reconstruction can very well be correlated with the same low-redshift structures used in the cross-correlation. For instance, in the case of cross-correlations with the galaxy overdensity $\delta_g$, the large-scale structure bispectrum with the tSZ Compton-$y$ field $\langle \delta_g yy \rangle$ contributes a very large bias \cite{vanengelen,Baxter2018}. The bispectrum appears since the quadratic lensing estimator $\langle TT \rangle$ contains some part of a $\langle yy \rangle$ contribution depending on the efficacy of foreground cleaning or whether any cleaning was done at all. This suggests that if even just one of the maps used is free of tSZ contamination, the bispectrum contribution to the bias will be zero.

\subsection{Existing foreground-cleaned maps}
\begin{figure}[t]
\includegraphics[width=8cm]{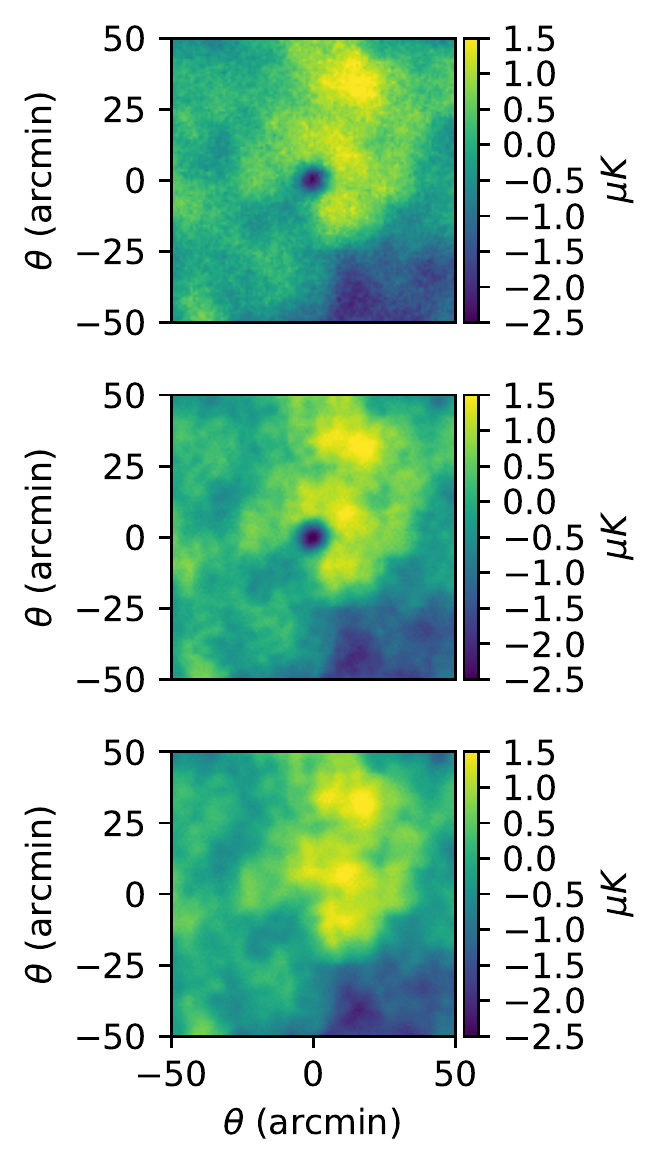}
\centering
\caption{Stacks on different CMB temperature maps at the locations of SDSS DR8 redMaPPer clusters. {\it Top:} The {\it Planck} 143 GHz map. The expected tSZ decrement at this frequency is clearly seen. {\it Center:} The {\it Planck} foreground-cleaned SMICA map. A large tSZ residual is visible, because the tSZ contamination is not explicitly deprojected in SMICA. {\it Bottom:} The LGMCA {\it Planck+WMAP} foreground-cleaned map. There is no visible evidence of any foreground residual in this stack.  In particular, LGMCA explicitly deprojects the tSZ foreground using its known frequency dependence.}\label{fig:stack}
\end{figure}

The {\it Planck} satellite has measured the CMB temperature and polarization in nine frequency bandpasses from 30 -- 857 GHz, while the earlier {\it WMAP} satellite made similar measurements in five frequency bandpasses from 23 -- 94 GHz. Various component-separation methods have been applied to these data sets to isolate the CMB signal while cleaning foreground contamination (see, e.g.,~\cite{PlanckFG2015,PlanckCompSep2015} for an overview of methods applied by the {\it Planck} team).

Here, we focus on two foreground-cleaned maps.  The first is a CMB map constructed from {\it Planck} data via the Spectral Matching Independent Component Analysis (SMICA) method~\cite{SMICAmethod}.  This method constructs a cleaned CMB map via a semi-blind internal linear combination in harmonic space.  Its noise properties are fairly similar to that of the Needlet ILC (NILC)~\cite{NILCmethod} CMB map produced by {\it Planck}, and thus we take it as a representative example here.  Importantly, neither the SMICA nor NILC {\it Planck} CMB maps impose a constraint to explicitly deproject any particularly contaminant, e.g., the tSZ effect.  Rather, the methods seek in general to solely minimize the total variance in the final map (while preserving the CMB signal).  The SMICA map forms the basis of the {\it Planck} 2015 CMB lensing reconstruction~\cite{plancklensing2015}.

The second map that we consider is a CMB map constructed from \pwmap data via the Local-Generalized Morphological Component Analysis (LGMCA) method~\cite{LGMCA1,LGMCA2}.  The underlying mathematical difference between LGMCA and the methods described above is that LGMCA imposes a sparsity criterion on the sources into which the data are decomposed.  LGMCA operates on a wavelet frame (as does NILC), where it finds the set of sources and mixing matrix that best fit the data, subject to a sparsity criterion on the sources' wavelet coefficients.

For our purposes, the most significant difference between LGMCA and SMICA (or NILC) is that the LGMCA CMB reconstruction imposes an explicit deprojection of the tSZ signal.  In other words, LGMCA enforces a constraint requiring the weights applied to the multifrequency maps to have zero response to the tSZ spectral function (which is known {\it a priori}).  Such a constraint can be applied in the ILC context as well~\cite{Remazeilles2011,Hillinprep}, but this was not done in producing the {\it Planck} 2015 SMICA and NILC CMB maps.  Finally, we note that although {\it WMAP} data is included in the LGMCA map, the information content in the final map is dominated by {\it Planck} on scales $\ell \gtrsim$ several hundred.

To demonstrate the importance of a map like LGMCA that explicitly deprojects the tSZ contamination, we stack on the locations of SDSS DR8 redMaPPer galaxy clusters~\cite{Redmapper}. In Figure~\ref{fig:stack}, the top panel shows a stack on the {\it Planck} 143 GHz temperature map. The expected tSZ decrement can be seen in this stack. In the middle panel, we show the same stack, but on the foreground-cleaned SMICA map. There is clear evidence of a large tSZ residual in this map, so it is not advisable to use the SMICA map to obtain cleaned CMB gradients. In the bottom panel, we show the stack on the LGMCA map. There is no evidence of any foreground residual in this stack.  In particular, the tSZ bias seen in the SMICA map is not seen here.  This demonstrates the successful deprojection of the tSZ signal in the LGMCA CMB reconstruction.

\begin{figure}[t]
\includegraphics[width=8cm]{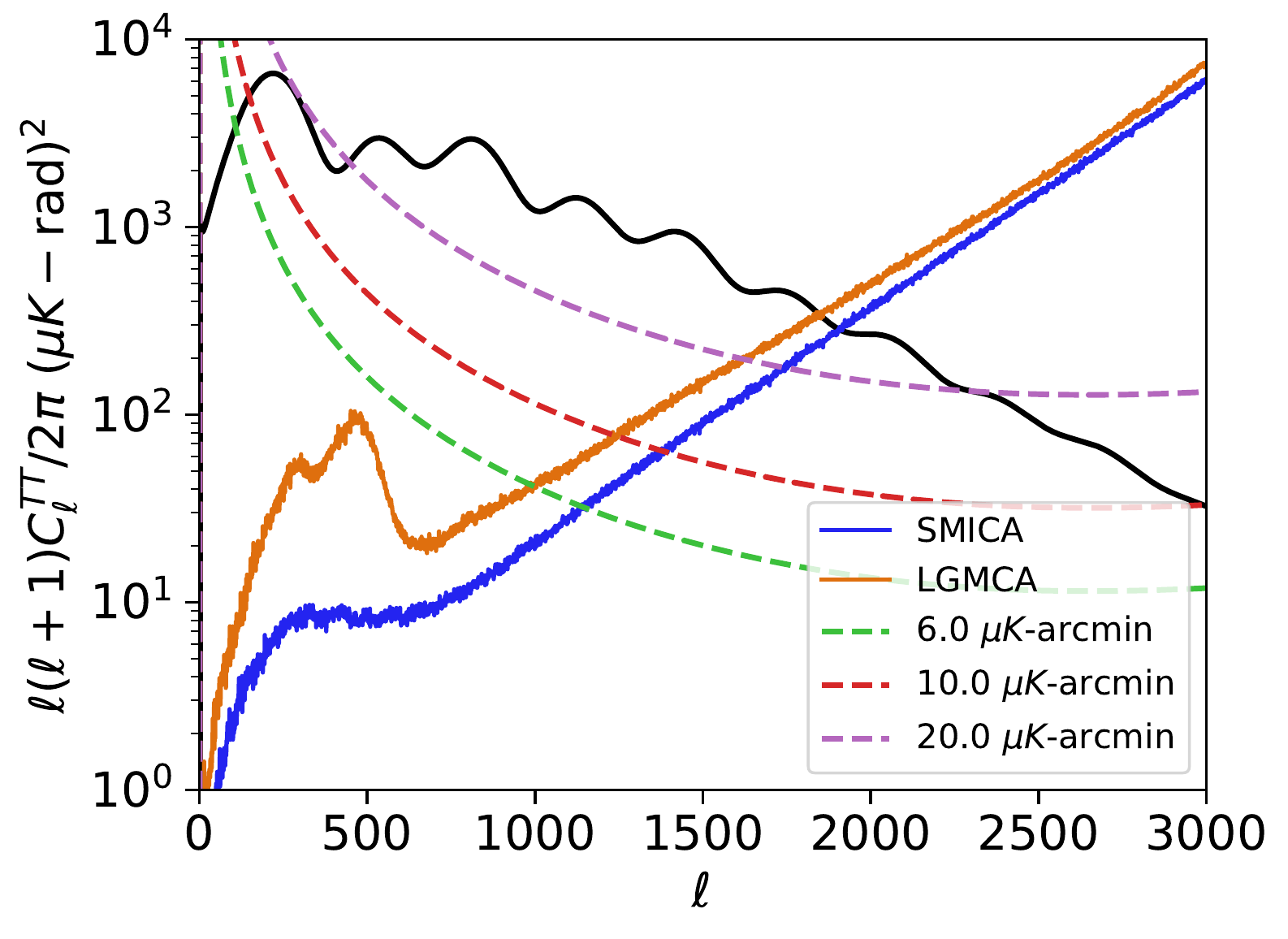}
\centering
\caption{Beam-deconvolved noise in CMB temperature maps. The orange solid curve shows the measured noise power spectrum in the LGMCA \pwmap foreground-cleaned maps. The blue solid curves shows the same for the {\it Planck} SMICA map. The dashed curves show the noise power spectra for various representative high-resolution, ground-based experiments with white noise levels of 6, 10, and 20 $\mu$K-arcmin, beam of FWHM = 1.5 arcminutes, and atmospheric noise with $\lknee=3000$ and $\alpha=-4$.  The black solid curve shows the fiducial CMB temperature power spectrum. }\label{fig:cmbNoise}
\end{figure}

In Figure~\ref{fig:cmbNoise}, we compare the beam-deconvolved noise power of the SMICA and LGMCA maps along with representative noise curves from ground-based, high-resolution experiments. The SMICA noise curve is obtained from a combination of auto-spectra and cross-spectra of ``half-ring'' splits of the {\it Planck} data. The LGMCA noise curve is obtained from the auto-spectrum of a provided noise map, also inferred from half-ring splits. The beam full-width-half maximum (FWHM) is 5 arcminutes for both maps. For the ground-based experiments, we assume instrumental white noise $s_{\nu,w}$ of 6, 10, or 20 $\mu$K-arcmin, effective beam of FWHM $\theta_{\rmn{FWHM}}= 1.5$ arcminutes, and an atmospheric contribution with $\ell_{\mathrm{knee}}=3000$ and $\alpha=-4$ roughly chosen to match existing measurements~\cite{Louis2017}. These parameters enter the beam-deconvolved noise power as

\begin{equation}
N_\nu(\ell) = s_{\nu,w}^2\left(1+\left(\frac{\ell}{\ell_{\mathrm{knee}}}\right)^{\alpha}\right)\rmn{exp}\left(\frac{\ell(\ell+1)\theta_{\rmn{FWHM}}^2}{8\rmn{ln}2}\right).
\end{equation}
The LGMCA and SMICA noise curves are fairly comparable above $\ell=1800$ where they start to become comparable to the CMB power, so lensing maps derived from LGMCA should not be significantly noisier than those based on SMICA.\footnote{However, note that the LGMCA noise power is higher than the SMICA noise power (by 22\% at $\ell=2000$), as it must be due to the use of one degree of freedom for the tSZ deprojection in LGMCA.}  The ground-based experiments do provide signal-dominated modes at scales $\ell>1500$, where the {\it Planck} maps become noisy, so some S/N penalty is expected. We quantify these statements in Section~\ref{sec:SN}.

\section{Simulations}\label{sec:Sims}

Having motivated cleaned CMB gradients for lensing reconstruction in the previous section, we now demonstrate using realistic simulations the existence of the tSZ bias and how it is eliminated when using cleaned gradients. We use simulations of the microwave sky from \cite{Sehgal2011}. From these simulations, we use the CMB lensing convergence map and a map of the tSZ Compton-$y$ field evaluated at 148 GHz.\footnote{We apply an additional scaling factor of 0.91 to the tSZ map to (approximately) match the tSZ power spectrum from the templates in \cite{bbps2010} and \cite{bbps2012b} used in \cite{Dunkley2013}.}

\subsection{CMB halo lensing}

We split the halo catalog from \cite{Sehgal2011} into four mass bins, the mass ranges of which are specified in Table~\ref{table:cluster-bins}. We only use halos at redshift $z>0.5$ since this is the sample for which CMB lensing will be primarily informative. For each object in the halo catalog, we simulate a lensing convergence stamp assuming an NFW profile with concentration of 3.2 and mass obtained from the halo catalog. We avoid using the lensing convergence from the simulations themselves so as to have control over the profiles we fit. We are not interested in testing cluster profiles or the contribution from the 2-halo term for this exercise; our interest is solely in the relative effect of tSZ contamination. However, we do use cutouts from the tSZ field at 148 GHz from the simulations centered on each halo from the halo catalog, as described below.

We generate periodic realizations of the unlensed CMB on 100 arcminute-wide postage stamps with a pixel width of 0.2 arcminutes. These are lensed using 5th order spline interpolation on displaced pixel positions with the above simulated NFW lensing convergence stamps. To the lensed CMB, we either add or do not add the tSZ stamp depending on the cases considered.

We convolve these sky simulations with a beam and add Gaussian white noise depending on the instrument combinations considered and then proceed to perform the asymmetric lensing reconstruction described in Section~\ref{sec:AsymLens}. All of these measurements are simulated at 148 GHz (this assumption only affects how we scale the tSZ fields from the simulations). However, our findings are generalizable to many kinds of correlated foregrounds at other frequencies (e.g., CIB). The reconstructions in each mass bin are stacked and the resulting stack is fit to an NFW profile parameterized only by mass. We finally compare the resulting mass estimates for various configurations.

\begin{figure}[t]
\includegraphics[width=8cm]{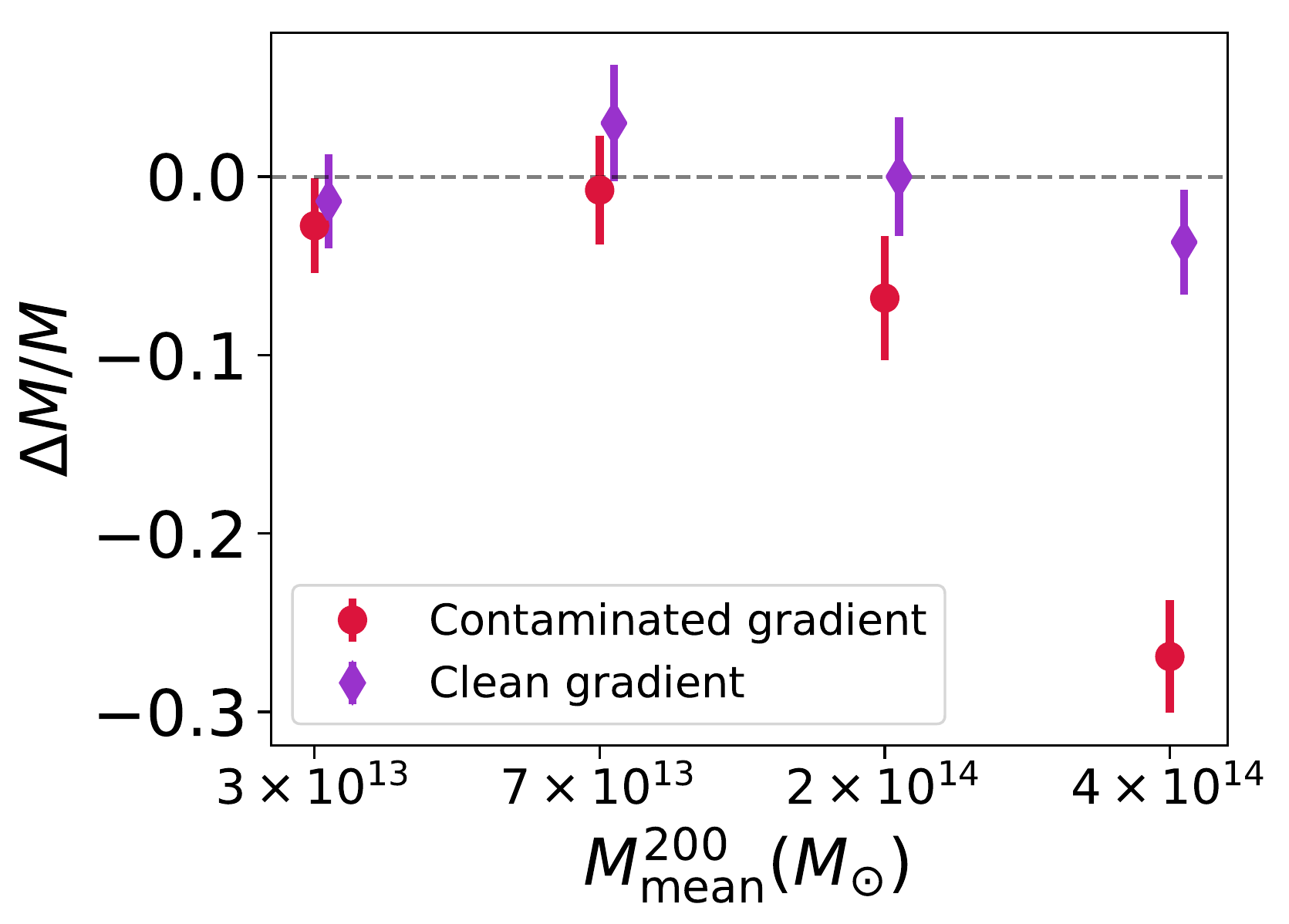}
\centering
\caption{Mass estimation bias for four different halo mass bins as determined from map-based simulations. The red circles show the relative mass bias when both the gradient and the high-resolution map are contaminated with tSZ signal (at levels expected at 150 GHz). For medium- and high-mass clusters, the bias is very large. The purple diamonds show the case when the gradient is free from tSZ contamination, but the high-resolution map contains tSZ. The bias is no longer detectable. The horizontal axis is neither on a linear nor logarithmic scale; the mass bin ranges on the horizontal axis are listed in Table~\ref{table:cluster-bins}.}\label{fig:clusterBias}
\end{figure}

Our experiment configuration for this test is as follows. We separately treat the gradient and high-resolution maps in the asymmetric quadratic estimator. The map that is used for the gradient is convolved with a FWHM = 5 arcminute beam and added to a realization of Gaussian white noise with amplitude $45$ $\mu$K-arcmin (similar to SMICA/LGMCA), while the high-resolution map uses a FWHM = 1.5 arcminute beam and white noise with amplitude $1.5$ $\mu$K-arcmin. We first perform TT estimator reconstructions when no tSZ has been added to either the gradient or the high-resolution map. All subsequent mass fit results are quoted with respect to this case. We then compare two cases: (i) both the gradient and high-resolution maps have tSZ in them; (ii) the gradient has no tSZ while the high-resolution map does. We show the results in Figure~\ref{fig:clusterBias}. While the two lowest mass bins are not affected (since the level of tSZ in them is small), the two highest mass bins receive significant negative biases from tSZ.  Physically, this bias appears for the following reason. Lensing by a circularly symmetric cluster causes a small scale dipole fluctuation anti-aligned with the background gradient. Thus, the quadratic estimator is normalized such that it interprets small-scale fluctuations anti-aligned with the background gradient as positive lensing convergence. A circularly symmetric foreground affects both the gradient estimate and the small-scale fluctuation estimate, but these are aligned in the same direction, and hence yield negative lensing convergence. The net effect is a scale-dependent suppression of the measured lensing convergence profile.  In contrast, in the case where the gradient is clean, our results are consistent with no tSZ bias.

\subsection{Large-scale lensing cross-correlation with halos}

We now perform a simulated cross-correlation analysis to demonstrate the effectiveness of gradient cleaning for applications to large-scale lensing cross-correlations. We divide an octant of the simulations from \cite{Sehgal2011} into 42 patches of area 100 deg${}^2$ each. Cutouts of the lensing convergence field and the tSZ field at 148 GHz are projected from \texttt{HEALPIX} to \texttt{CAR} pixelization with a pixel size of 0.5 arcminutes on to these patches. These are then resampled through Fourier-space trimming down to a pixel size of 2 arcminutes.  For each of the 42 patches, we generate 200 realizations of the unlensed CMB and then lens these with the cut-out lensing convergence map using the same algorithm as for the halo lensing case. Once again, the map that is used for the gradient is convolved with a FWHM = 5 arcminute beam and added to a realization of Gaussian white noise with amplitude $45$ $\mu$K-arcmin, while the high-resolution map uses a FWHM = 1.5 arcminute beam and white noise with amplitude $10$ $\mu$K-arcmin. Later, we consider cases where the tSZ  signal has either been added or not added to the lensed maps used in either leg of the quadratic estimator (before beam convolution). We also construct a halo overdensity map by taking all halos of mass $M_{200} >1 \times 10^{13} \, M_{\odot}$ from the halo catalog with redshift $0.2<z<0.8$ and  making a histogram of their positions, where the histogram bins are the pixels in our map geometry. This results in a count map $n_h(\bth)$ from which we obtain the ``galaxy overdensity'' used for cross-correlation as follows:\footnote{Of course, we have not populated these halos with galaxies in any biasing framework, but this detail is irrelevant to the main conclusion.}
\begin{equation}
\delta_h(\bth)  = \frac{n_h(\bth)}{\bar{n}_h}-1 \,,
\end{equation}
where $\bar{n}_h$ is the mean number of halos in our sample and $\bth$ is the position of a pixel.

\begin{figure}[t]
\includegraphics[width=8cm]{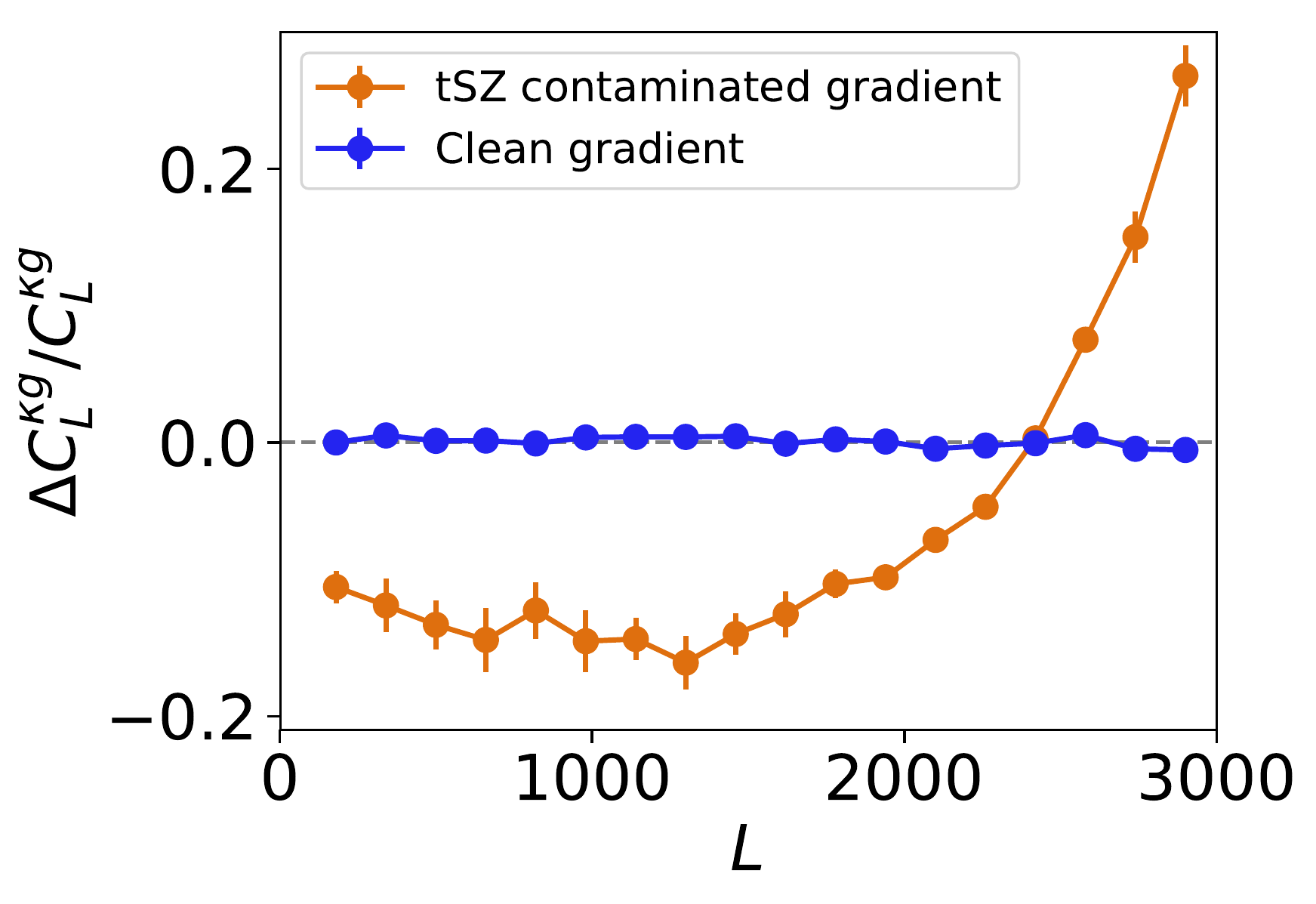}
\centering
\caption{The tSZ bias to the CMB lensing-galaxy density cross-correlation as measured from simulations, with and without clean gradients. Both curves here are relative differences with respect to the case where no tSZ was added to the CMB maps. We show the case where tSZ contamination is present in both the gradient and high-resolution CMB maps (orange) and the case where the tSZ contamination is only in the high-resolution map and the gradient does not contain tSZ (blue). The error bars show the uncertainty on our estimate of these biases when averaged across 42 patches of 100 deg${}^2$ each.}\label{fig:crossBias}
\end{figure}

Once again, we compare two reconstruction scenarios to the case where no tSZ contamination is added at all: (i) tSZ in both the gradient and high-resolution map; (ii) no tSZ in the gradient. In any given one of the 42 patches, we reconstruct lensing for the 200 CMB realizations using the asymmetric TT estimator for these two cases and cross-correlate these reconstructed lensing maps with the galaxy overdensity map $\delta_h(\bth)$ to obtain the cross-power-spectra $C_L^{\kappa g}$ in annular bins in multipole space. We find the mean across 200 simulations of these bandpowers in each patch (to average over instrumental and CMB noise) and calculate the relative difference with respect to the case in which no tSZ is added to either of the CMB maps. We then calculate the mean and covariance of these relative differences across the 42 patches to capture the contribution to uncertainties from sample variance.  The mean relative differences are shown in Figure~\ref{fig:crossBias}. Once again, when the CMB gradient is contaminated, there is a large bias at both large and small scales, but this bias disappears when the gradient alone is clean. The bias is as large as 19\% for our halo selection, but we note that the level of bias can be larger and is sensitive to the minimum mass and redshift range of halos selected (as also seen in \cite{Baxter2018}). With clean gradients, any bias is well below the 1\% level.

\begin{table}[]
\centering
\caption{The mass ranges, mean redshifts, and number of halos for each bin used in lensing reconstruction from simulated halos with and without tSZ contamination. Note that only halos at $z>0.5$ are considered in our analysis.}
\label{table:cluster-bins}
\begin{tabular}{l|lllll}
                & \textbf{\begin{tabular}[c]{@{}l@{}}$M^{200}_{\mathrm{min}}$\\ ($M_{\odot}$)\end{tabular}} & \textbf{$M^{200}_{\mathrm{max}}$($M_{\odot}$)} & \textbf{\begin{tabular}[c]{@{}l@{}}$M^{200}_{\mathrm{mean}}$($M_{\odot}$)\end{tabular}}  & \textbf{$z_{\mathrm{mean}}$} & \textbf{$N$} \\ \hline
\textbf{Groups} & $1 \times 10^{13}$& $5 \times 10^{13}$& $3.0\times 10^{13}$ & 0.53 & 30000        \\
\textbf{Low}    & $5 \times 10^{13}$& $1 \times 10^{14}$& $6.8\times 10^{13}$ & 0.53 & 10000        \\
\textbf{Medium} & $1 \times 10^{14}$& $3 \times 10^{14}$& $1.5\times 10^{14}$ & 0.54 & 5000         \\
\textbf{High}   & $3 \times 10^{14}$ & - & $4.2\times 10^{14}$ & 0.77 & 2500        
\end{tabular}
\end{table}

\section{Cross-correlation signal-to-noise with cleaned gradients} \label{sec:SN}

How is the S/N for cross-correlations affected if we replace the gradient of the quadratic estimator for TT with the LGMCA foreground-cleaned (tSZ-free) map? We want to compare lensing reconstructions involving $\mathrm{QE}( [\nabla T]_{\mathrm{LGMCA}} , [T]_{\mathrm{Hi-res}} )$ and $\mathrm{QE}( [\nabla T]_{\mathrm{Hi-res}} , [T]_{\mathrm{Hi-res}} )$. Instead of directly comparing  the performance of these two estimators, we compare the more realistic scenario where polarization data is available as well, by constructing a minimum-variance combination of TT, ET, EE, EB, and TB, where in the case involving LGMCA, the T map used in the gradient of TT is assumed to be from LGMCA. The high-resolution experiment is also assumed to have atmospheric noise as described in Section~\ref{sec:ForeBias} and shown in Figure~\ref{fig:cmbNoise}. In addition, the minimum CMB multipole used is assumed to be $\ell_{\mathrm{min}}=500$ for the high-resolution experiment and $\ell_{\mathrm{min}}=2$ for LGMCA. For all cases, we use $\ell_{\mathrm{max}}=3000$ for temperature and $\ell_{\mathrm{max}}=5000$ for polarization.

\begin{figure}[t]
\includegraphics[width=8cm]{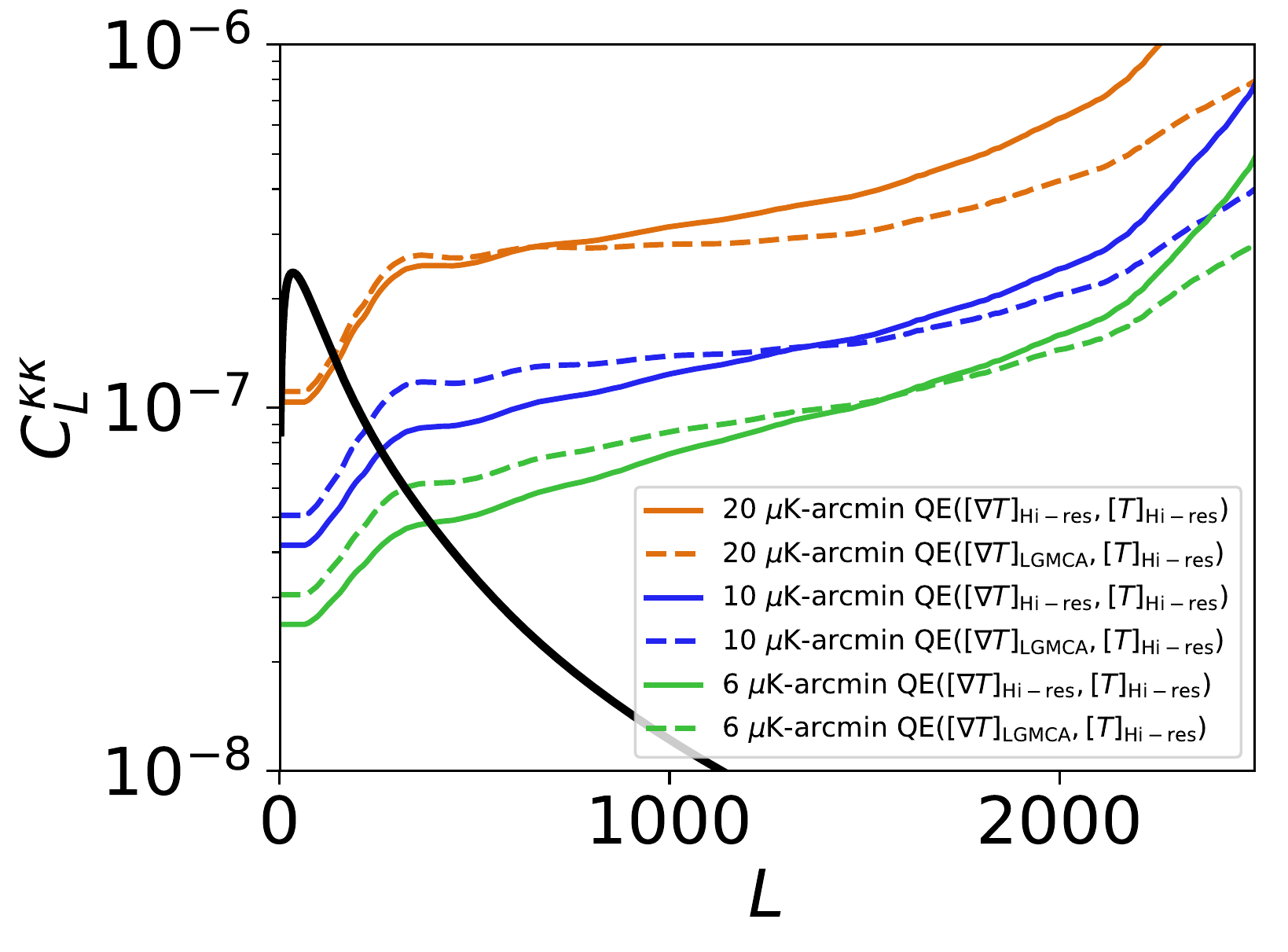}
\centering
\caption{Lensing noise curves for asymmetric lensing reconstruction compared to symmetric reconstruction.  The solid curves show the lensing noise power spectrum when the high-resolution CMB experiment is used in both legs of the quadratic estimator for TT. This case is expected to have a very significant tSZ bias. The dashed curves show the case when the temperature gradient leg of the TT estimator is replaced with the LGMCA (tSZ-free) map. This case has slightly lower S/N at large scales, but is expected to be free of tSZ bias. All curves are minimum-variance combinations of TT, ET, EB, EE and TB.}\label{fig:lensSn}
\end{figure}

In Figure~\ref{fig:lensSn}, we show the resulting lensing noise curves calculated from the quadratic estimator normalization as $N_L^{\kappa\kappa}=L^2A_L/4$. As expected, the lensing noise curves are slightly higher on large scales due to the gradient measurement in TT being noisier. For small scales, the reconstruction involving LGMCA does better. This is almost entirely due to the choice of $\ell_{\mathrm{min}}=2$ for LGMCA compared to $\ell_{\mathrm{min}}=500$ for the high-resolution experiment, and partially due to atmospheric noise; as noted earlier, the small-scale lens squeezed limit gets most of its S/N from the correlation of large-scale gradients and small-scale fluctuations.

In Table~\ref{table:sn}, we compile these findings into S/N forecasts for various cross-correlation measurements. We forecast for CMB lensing cross-correlated with galaxy lensing, and for CMB lensing cross-correlated with galaxy overdensity. We make assumptions about optical surveys that are not intended to be extremely precise, but only roughly representative, allowing for a comparison of the effect of using LGMCA gradients. The optical surveys we consider are the Dark Energy Survey (DES) \cite{des} and the Hyper Suprime-Cam (HSC) survey \cite{hsc}. Similar conclusions should also hold for the Kilo-Degree Survey (KiDS)  \cite{kids} and future surveys like LSST \cite{lsst}, WFIRST \cite{wfirst}, and Euclid \cite{euclid}. For both DES and HSC, we assume a photometric redshift distribution given by

\begin{equation}
\frac{dN}{dz}(z) = \frac{z^2}{2z_0^3} \rmn{exp}\left(-\frac{z}{z_0}\right)  
\end{equation}
with $z_0=1/3$. We assume the same galaxy number density for the galaxy overdensity and galaxy lensing forecast, with $n_{\rmn{gal}}=6\ \rmn{arcmin^{-2}}$ for DES and $n_{\rmn{gal}}=30\ \rmn{arcmin^{-2}}$ for HSC. We assume that the CMB lensing survey overlaps with DES over 5000 square degrees and with HSC over 1400 square degrees. The S/N is calculated assuming a Gaussian distribution for the bandpowers as 

\begin{equation} \label{eq:sngal}
\begin{aligned}
\left( \frac{S}{N} \right)^2 &= f_{\mathrm{sky}} \, \times  \\ 
& \sum_{L=L_{\min}}^{L_{\max}} \frac{(2 L + 1) \left(C_L^{\kappa X} \right)^2}{\left(C_L^{\kappa X} \right)^2 + \left(C_L^{\kappa \kappa} + N_L^{\kappa \kappa}  \right) \left(C
_L^{XX} + N_L^{XX} \right) }                 
\end{aligned}
\end{equation} 
where $X=g$ for the galaxy overdensity and $X=\gamma$ for the galaxy lensing shear. We include noise from the galaxy surveys as $N_L^{gg}=1/n_{\rmn{gal}}$ and $N_L^{\gamma\gamma}=\sigma^2_s/(2 n_{\rmn{gal}})$ where the shape noise is assumed to be $\sigma_s=0.3$. 

As seen in Table~\ref{table:sn}, we find that for $20$ $\mu$K-arcmin noise for the high-resolution survey, there is a slight improvement in using LGMCA gradients over using gradients from the high-resolution experiment due to the choice of $\ell_{\rmn{min}}$ mentioned earlier. At lower noise levels, the polarization estimators prevent the S/N from degrading significantly. We find that the S/N penalty in using LGMCA temperature gradients is never greater than 5.4\% if polarization data is used in conjunction with temperature. If the TT estimator alone is considered, then the S/N penalty in using LGMCA temperature gradients is never greater than 14.5\%.

\begin{table*}[]
\centering
\caption{The S/N for various CMB lensing cross-correlations. We compare the case where the noisier LGMCA \pwmap map is used for the gradient of the temperature in the TT estimator with the standard case of using the high-resolution (potentially contaminated) CMB maps in both legs. Due to the presence of atmospheric noise and a conservative choice of $\ell_{\rmn{min}}=500$ for the ground-based experiment, the LGMCA case is slightly better for $20$ $\mu$K-arcmin noise. For lower noise levels, the penalty in S/N is small ($<5.4\%$) while allowing for tSZ-bias-free measurements. All forecasts involve minimum-variance combinations of all the temperature and polarization quadratic estimator combinations.}
\label{table:sn}
\begin{tabular}{l|llll}
                                                                                                 & \textbf{DES $C_{\ell}^{\kappa \gamma}$} & \textbf{DES $C_{\ell}^{\kappa g}$} & \textbf{HSC $C_{\ell}^{\kappa \gamma}$} & \textbf{HSC $C_{\ell}^{\kappa g}$} \\ \hline
\textbf{20 $\mu$K-arcmin } &&&&\\
\textbf{$\mathrm{QE}( [\nabla T]_{\mathrm{Hi-res}} , [T]_{\mathrm{Hi-res}} )$} &37&70&24&38\\
\textbf{$\mathrm{QE}( [\nabla T]_{\mathrm{LGMCA}} , [T]_{\mathrm{Hi-res}} )$} &37&72&24&41\\
S/N difference &+1.6\%&+3.4\%&+2.8\%&+3.8\%\\
\textbf{10 $\mu$K-arcmin } &&&&\\
\textbf{$\mathrm{QE}( [\nabla T]_{\mathrm{Hi-res}} , [T]_{\mathrm{Hi-res}} )$} &53&102&35&56\\
\textbf{$\mathrm{QE}( [\nabla T]_{\mathrm{LGMCA}} , [T]_{\mathrm{Hi-res}} )$} &50&98&33&54\\
S/N difference &-5.4\%&-4.3\%&-4.8\%&-4.0\%\\
\textbf{6 $\mu$K-arcmin } &&&&\\
\textbf{$\mathrm{QE}( [\nabla T]_{\mathrm{Hi-res}} , [T]_{\mathrm{Hi-res}} )$} &63&122&42&68\\
\textbf{$\mathrm{QE}( [\nabla T]_{\mathrm{LGMCA}} , [T]_{\mathrm{Hi-res}} )$} &60&117&40&65\\
S/N difference &-4.7\%&-4.1\%&-4.5\%&-4.0\%\\
    
\end{tabular}
\end{table*}

\section{Discussion and Conclusion}\label{sec:Conclusion}

Astrophysical foregrounds, such as the tSZ effect, can introduce large biases in the cross-correlation of CMB lensing with low-redshift tracers and in the inference of halo masses using CMB lensing. In this work, we have shown that these biases can be eliminated using cleaned CMB gradients obtained from existing data from the {\it Planck} and {\it WMAP} experiments.  

While our conclusions can be generalized to other foreground contaminants like the CIB, we specifically advocate the use of a tSZ-deprojected map like the LGMCA map for CMB cluster lensing and cross-correlation measurements, since tSZ is expected to be the dominant contaminant for these measurements.\footnote{Deprojection of an assumed CIB-like spectrum in CMB component separation should thus be expected to remove similar biases for cross-correlations with high-redshift samples, e.g., quasars.  Multi-component deprojection~\cite{Hillinprep} could be used for maxium robustness, with a slightly larger S/N penalty.} We have shown that using the LGMCA \pwmap map for CMB gradients in place of potentially contaminated high-resolution measurements does not degrade the S/N of lensing cross-correlations by more than 5.4\%. 

The advantages of this approach are very significant. In \cite{Baxter2018}, the authors show how worrying the tSZ-induced bias in CMB lensing cross-correlations can be. That work advocates aggressive masking of tSZ clusters and imposing additional angular scale cuts on the cross-correlation to be less sensitive to the bias. Aggressive masking affects the interpretation of the cross-correlation measurement since regions of high mass density are being masked (i.e., the mask is not unbiased with respect to the field that is being measured). Estimates of whether or not the level of masking affect the interpretation then become tied to the accuracy of the simulations used in estimating the size of the bias. In \cite{Baxter2018}, the authors estimate the size of the bias in a way that is partially informed by data: by cross-correlating their galaxy sample with tSZ signal painted in on the location of known redMaPPer clusters in the survey region. This approach must underestimate the bias by some amount: as the authors mention, there are contributions from halos below the redMaPPer detection threshold and redshift range, as well as diffuse tSZ outside clusters. Moreover, mitigating these biases through aggressive masking and scale cuts can increase the uncertainty on parameters by a factor of 2 \cite{Baxter2018}. Gradient cleaning in combination with polarization data will only increase the uncertainty on a parameter like $\sigma_8$ by at most 2.5\%. For these reasons, it becomes ever more compelling to use gradients from the LGMCA map (or other cleaned maps from {\it Planck} data that explicitly deproject the tSZ contamination). No masking, in-painting, or angular scale cuts are necessary, practically no S/N penalty is incurred, and the interpretation of the resulting measurement becomes much simpler.

The bias to CMB lensing cross-correlations shown in Figure~\ref{fig:crossBias} becomes more significant at larger scales. Some recent cross-correlation measurements of CMB lensing with low-redshift galaxy overdensities have in fact exhibited deficits with respect to theoretical predictions on large scales. In \cite{pullenho}, the authors cross-correlated the {\it Planck} CMB lensing map (which was derived from the tSZ-contaminated SMICA map) with BOSS spectroscopic galaxies ($z_{\rmn{mean}}=0.57$) and found that the inferred $E_G$ statistic\footnote{See, e.g., \cite{egpullen} for a definition of $E_G$. This statistic is formed from a combination of the lensing-galaxy cross-correlation, galaxy auto-correlation, and the redshift-space distortion parameter, and is a probe of deviations from General Relativity.} was in 2.6$\sigma$ tension with $\Lambda CDM$+GR. This tension was driven by a large-scale deficit in the cross-correlation.  In~\cite{HillkSZ2,FerrarokSZ2}, a sample of infrared-selected galaxies from {\it WISE} was cross-correlated with the {\it Planck} CMB lensing map in order to calibrate the {\it WISE} galaxies' bias.  A similar, anomalously low cross-correlation at $L<100$ was observed, and subsequently this multipole range was not used in the analysis.  The deficit was present when considering both the 2013 and 2015 {\it Planck} CMB lensing maps.  Similarly, the cross-correlation measurement between CMB lensing and DES galaxies in \cite{Giannantonio} also showed a 1.7$\sigma$ deficit. The CMB lensing maps used there were derived from SPT-SZ observations at 150 GHz and the {\it Planck} SMICA map, both of which contain tSZ contamination. A re-analysis of these measurements using clean gradients from LGMCA or other tSZ-free maps could shed light on these tensions.  It would also be useful to explicitly compute CIB- or kSZ-related lensing biases for these measurements as well.

The analysis presented here has been based on the quadratic estimator since it naturally allows for a separate treatment of the CMB gradient and small-scale fluctuations. However, the quadratic estimator is known to be both biased and sub-optimal when applied to small-scale lenses like clusters. The bias (that appears even in the absence of foregrounds) is mass-dependent, and is due to regions of high lensing convergence breaking the gradient approximation that the estimator is derived under. This bias only becomes appreciable for the most massive halo bins and can easily be calibrated out using simulations. For mass measurements approaching percent-level precision, alternate techniques will be required. The sub-optimality at small scales for the quadratic estimator becomes important for noise levels less than around 4 $\mu$K-arcmin. Maximum-likelihood techniques are promising in this regard for measurements that are free of mass-dependent bias and optimal~\cite{Baxter2015,Raghu}. However, these can be highly susceptible to foregrounds~\cite{Raghu}, and further work is needed to incorporate proper marginalization over foregrounds while ensuring that the computational requirements do not make the approach infeasible. One promising method is the gradient inversion technique \cite{gradInversion} since it allows for an asymmetric treatment of the gradient as advocated here. Another approach is to iteratively delens the CMB map with quadratic estimator reconstructions~\cite{yooz1,yooz2}. This should in theory be equivalent to maximum-likelihood, and an investigation of this technique in the presence of foregrounds (taking advantage of clean gradients) is left for future work. 

Gradient cleaning can also help measurements of the CMB lensing auto-spectrum $C_L^{\kappa\kappa}$. The relevant observable here is the trispectrum or 4-point function of the CMB, e.g., $\langle T(\bEll_1)T(\bEll_2)T(\bEll_3)T(\bEll_4) \rangle$. To measure $C_L^{\kappa\kappa}$ from this 4-point function, we hope to isolate the connected component. For example, for TT in the asymmetric formalism, schematically we have

\begin{equation}
\begin{aligned}
\langle \boldsymbol{\nabla}TT\boldsymbol{\nabla}TT \rangle = & \langle \boldsymbol{\nabla}TT\rangle\langle\boldsymbol{\nabla}TT \rangle + \langle \boldsymbol{\nabla}TT\rangle\langle\boldsymbol{\nabla}TT \rangle  \\
& + \langle \boldsymbol{\nabla}T\boldsymbol{\nabla}T\rangle\langle TT \rangle + \langle \boldsymbol{\nabla}TT\boldsymbol{\nabla}TT \rangle_c  
\end{aligned}
\end{equation}
where the subscript $c$ denotes the connected component encoding the lensing induced non-Gaussianity. One typically subtracts the disconnected part (the $N_0$ bias) from simulations, and this subtraction is done using combinations that involve the data such that the subtraction is robust to mismatches between power in the data and power in the simulations \cite{namikawa}. It is clear that if the gradient above is free from foregrounds, we obtain no bias from the intrinsic connected component of the potentially non-Gaussian foregrounds, since foregrounds are not expected to form connected 4-point functions in combination with the lensed CMB. In Appendix A, we show that the usual realization-dependent $N_0$ subtraction \cite{namikawa} also absorbs any additional disconnected contribution, thus ensuring that the entire measurement is free from foreground bias with only cleaning of the gradients required. A detailed investigation of gradient cleaning in the context of autospectra involving tests on simulations and comparisons of S/N is beyond the scope of this paper; we leave it for future work.  (For the configurations considered in Table \ref{table:sn}, the largest S/N penalty for $C_L^{\kappa\kappa}$ we find is 12.7\%.)  We also note that foreground biases due to the kSZ effect cannot be mitigated by any type of multi-frequency component separation (although they are expected to be smaller than those due to tSZ contamination), and will thus require different strategies than those considered here~\cite{kszbias}.

Ongoing and planned CMB experiments like Advanced ACT \cite{advactpol}, SPT-3G \cite{spt3g}, Simons Array \cite{simonsarray1,simonsarray2}, Simons Observatory, and CMB-S4 \cite{cmbs4} will improve on measurements made by {\it Planck} across the frequency spectrum. These experiments will then provide clean measurements of the gradient that are independent of {\it Planck}. Explicit deprojection of biasing foregrounds in a constrained ILC analysis will likely degrade the S/N at small scales, but we expect that requiring CILC only in the gradients will allow for robust, tSZ-bias-free (and CIB-bias-free) CMB lensing measurements with effectively no S/N penalty.

\acknowledgments

We thank David Spergel for suggesting this line of exploration. We also thank Nicholas Battaglia, Jo Dunkley, Neelima Sehgal, Blake Sherwin, and Alexander van Engelen for useful conversations, and Nam Ho Nguyen for valuable contributions to some of the code libraries used. JCH acknowledges support from the Friends of the Institute for Advanced Study.

\bibliographystyle{apj}
\bibliography{msm}

\begin{appendix}

\section{Gradient Cleaning for CMB Lensing Autospectra Foreground Bias}

Here, we show that if a measurement of the CMB gradient is available that is uncontaminated by astrophysical foregrounds, then it can be used in conjunction with possibly contaminated high-resolution CMB measurements without leading to a bias in the lensing auto-spectrum $C_L^{\kappa\kappa}$. While this paper has focused on temperature, we now switch attention to polarization. The first motivation for this is that the EB estimator will dominate the S/N for future experiments when map sensitivities are lower than around $4$ $\mu$K-arcmin. Galactic foregrounds are known to be polarized and so it is important to ensure that this introduces no bias in CMB lensing maps which will be used for both delensing of B-modes (and hence affect estimates of the tensor-to-scalar ratio $r$) and for measuring $C_L^{\kappa\kappa}$. It is clear from the arguments in the rest of this paper that cleaned gradients remove foreground-induced biases in delensing applications. (Whether this statement holds true in the case of iterative delensing is not clear and needs to be investigated.) The second motivation for considering the EB estimator is that it  allows for much clearer notation in this appendix.

Let us assume only the $B$-map is contaminated by foregrounds,
\begin{equation}
B^c = B+B^f \,,
\end{equation}
where $B$ is the lensed CMB signal and $B^f$ is the foreground B-mode contaminant.

In the following, we suppress all Fourier transforms and lensing filters and present the calculation schematically. The na\"{i}ve lensing power spectrum before removal of the Gaussian disconnected $N_0$ bias is

\begin{equation}
\begin{aligned}
\hat{C}^{\phi\phi} = & \langle EB^cEB^c \rangle \\
= & \langle EBEB \rangle + \langle EB^fEB^f \rangle \\
& + \langle EB^fEB \rangle + \langle EBEB^f \rangle 
\end{aligned}
\end{equation}

The first term contains the connected component that depends on $C_L^{\kappa\kappa}$  plus the Gaussian $N_0$ bias that is usually removed by a realization-dependent Monte-Carlo procedure, all in the case when foregrounds are absent. We next assume that foregrounds neither form connected components with the lensed CMB modes (appropriate for polarized Galactic foregrounds), nor do they correlate with the lensed CMB modes. Using Wick's theorem, this leaves only $\langle EE \rangle \langle B^fB^f \rangle$ as a bias to the lensing power spectrum, on top of the usual Gaussian bias of the uncontaminated maps:

\begin{equation}\label{eq:biasedAuto}
\hat{C}^{\phi\phi} = \langle EBEB \rangle + \langle EE \rangle \langle B^fB^f \rangle
\end{equation}

The realization-dependent bias subtraction is \cite{namikawa,sherwin2015}:

\begin{equation}
\begin{aligned}
\Delta C_{RDN0} = & \langle\langle EB^sEB^s \rangle + \langle E^sB^cEB^s \rangle \\
& + \langle E^sB^cE^sB^c \rangle + \langle EB^sE^sB^c \rangle \\
& - \langle E^sB^{s'}E^sB^{s'} \rangle - \langle E^sB^{s'}E^{s'}B^s \rangle\rangle_{s,s'}
\end{aligned}
\end{equation}
where $s$ and $s'$ are a simulated pair of CMB maps that are averaged over. Some of these terms cancel the Gaussian bias in the uncontaminated first term of $\hat{C}^{\phi\phi}$ above. The remaining terms expand to:
\begin{equation}
\begin{aligned}
\Delta C_{RDN0}^{\mathrm{res}} = & \langle\langle E^sB^fEB^s \rangle + \langle E^sB^fE^sB^f \rangle \\
& + \langle E^sB^fE^sB \rangle + \langle E^sBE^sB^f \rangle \\
& + \langle EB^sE^sB^f \rangle\rangle_{s}
\end{aligned}
\end{equation}

The same assumptions above about the foregrounds on the sky hold for our simulation-data combination, simplifying this to

\begin{equation}
\Delta C_{RDN0}^{\mathrm{res}} = \langle E^sE^s \rangle \langle B^fB^f \rangle \,,
\end{equation}
which absorbs the only remaining bias from foregrounds in the lensing power spectrum in Eq. \ref{eq:biasedAuto}, to the extent that the simulation power spectra $\langle E^sE^s\rangle$ match that of the data $\langle EE\rangle$.

\end{appendix}
\end{document}

%% file: mydefinitions.tex
\newcommand{\be}{\begin{equation}}
\newcommand{\ee}{\end{equation}}

